\documentclass[aps,superscriptaddress,amsfonts,amssymb,amsmath,amsmath,nofootinbib,twocolumn]{revtex4}
\usepackage[utf8]{inputenc}
 \usepackage[T2A]{fontenc}
 \usepackage[english]{babel}
\usepackage{graphicx,color}
\usepackage{tabularx}
\usepackage{comment}
\definecolor{darkblue}{rgb}{0,0,0.7}
\definecolor{darkred}{rgb}{0.7,0,0}
\definecolor{darkgreen}{rgb}{0.1,0.4,0.1}
\usepackage[unicode, colorlinks, citecolor=darkblue, linkcolor=darkred, urlcolor=blue]{hyperref}

\usepackage{euler}

\allowdisplaybreaks

\begin{document}

\title{Thermal charge carrier driven noise in transmissive semiconductor optics}

\author{F.~Bruns}
\affiliation{Technische Universit\"at Braunschweig, LENA Laboratory for Emerging Nanometrology, Pockelsstra{\ss}e 14, 38106 Braunschweig, Germany}

\author{S.~P.~Vyatchanin}
\affiliation{Faculty of Physics, M.V. Lomonosov Moscow State University, Moscow 119991, Russia,\\
Quantum Technology Centre, M.V. Lomonosov, Moscow State University, Moscow 119991, Russia}

\author{J.~Dickmann}
\affiliation{Physikalisch-Technische Bundesanstalt, Bundesallee 100, 38116 Braunschweig, Germany}

\author{R.~Glaser}
\affiliation{Institut f{\"u}r Festk{\"o}rperphysik, Friedrich-Schiller-Universit{\"a}t Jena, 07743 Jena, Germany}

\author{D.~Heinert}
\affiliation{Institut f{\"u}r Festk{\"o}rperphysik, Friedrich-Schiller-Universit{\"a}t Jena, 07743 Jena, Germany}

\author{R.~Nawrodt}
\affiliation{5. Physikalisches Institut, Universit{\"a}t Stuttgart, 70569 Stuttgart, Germany}

\author{S.~Kroker}
\affiliation{Physikalisch-Technische Bundesanstalt, Bundesallee 100, 38116 Braunschweig, Germany}
\affiliation{Technische Universit\"at Braunschweig, LENA Laboratory for Emerging Nanometrology, Pockelsstra{\ss}e 14, 38106 Braunschweig, Germany}

\date{ \today}

\begin{abstract}
 
Several sources of noise limit the sensitivity of current gravitational wave detectors. Currently, dominant noise sources include quantum noise and thermal Brownian noise, but future detectors will also be limited by other thermal noise channels. In this paper we study a thermal noise source which is caused by spatial charge carrier density variations in semiconductor materials. We provide an analytical model for the understanding of charge carrier fluctuations under the presence of screening effects and show that charge carrier noise will not be a limiting noise source for third generation gravitational wave detectors.

\end{abstract}

\maketitle

\section{Introduction}
Gravitational wave detection (GWD) \cite{GW_2016, GW_2016b} is an exciting and growing field in the realm of high-precision metrology. For the first detection of a gravitational wave, complex infrastructure was built taking huge care to reduce all of the possible noise sources. In particular the fascinating accuracy of a displacement measurement of a macroscopic test mass is demonstrated on a level of about $\sim 10^{-18}$ m~\cite{aLIGO2013,aLIGO2014,aLIGO2015,aLIGO2015b,aVirgo2015,aVirgo2018}. 

A third generation of gravitational wave detectors is planned in order to access gravitational waves as a new channel of information into the universe addressing fundamental questions such as the nature of dark matter, the inner structure of black holes and the origin of the universe itself~\cite{Etscience}. For third generation gravitational wave detectors (Einstein Telescope in Europe or Voyager and Cosmic Explorer in USA) the use of crystalline materials is considered~\cite{3Gen2018}. Third generation gravitational wave detectors are severely limited by thermal noise. Therefore there is a high probability they will be run at cryogenic temperatures. At these temperatures crystalline materials have superior mechanical quality factors compared to conventional amorphous materials thus being less susceptible to thermal Brownian noise~\cite{nawrodt2011,martin2010,martin2008,nawrodt2008,martin2014,nawrodt2007}. Therefore, semiconductors such silicon are considered as possible substrate materials. This class of materials is potentially susceptible to novel noise channels.

In this paper we study thermal carrier noise in semiconductor materials. Similary to thermochemical noise~\cite{benthem2009} which is caused by the diffusion of optical impurities this source of noise originates from the thermal motion of free charge carriers in the transmissive optical elements. The Brownian motion of charge carriers leads to spatial variations of the free carrier density thus creating local fluctuations of the refractive index. These fluctuations will be probed by any beam of light transmitted through the optical element. Below  we refer to this noise as thermal charge carrier refractive (TCCR) noise. For free charge carriers screening effects have to be considered which distinguishes TCCR noise from thermochemical noise. A quantitative analysis of TCCR noise is presented taking into account Debye screening which depresses TCCR noise.

\section{Problem statement}\label{problemstatement}
Earthbound gravitational wave detectors consist of a large scale Michelson interferometer with Fabry-Perot (FP) cavities in the arms for signal enhancement~\cite{bookSaulson}. An incoming gravitational wave with the strain $h$ displaces the end mirrors of the FP cavities by a length of $\delta l$:
\begin{align}\label{defstrain}
\delta l =\frac{L_0 h}{2}.
\end{align}
In order to measure the variation of the gravitational metric $h$ the displacement of the end mirrors is probed with a laser beam. The displacement of the end mirrors causes a phase shift between the interferometer arms which is enhanced by the FP-cavities and read out via interference of laser light of the two arm cavities.

In this paper we investigate transmissive optics made of semiconductor materials which is particularly important for third generation gravitational wave detectors such as the Einstein telescope. In semiconductor optics such as silicon~\cite{ETdesign} free charge carriers are present. These charge carriers move thermally akin to Brownian motion thus creating local charge carrier density fluctuations $\eta$. As stated by Soref \textit{et al.}~\cite{soref1987} the refractive index $n$ is dependent on the charge carrier concentration as an effect of the free carrier dispersion:
\begin{equation}
\delta n = \beta\, \eta,\quad \beta = \frac{-e^2}{2 \epsilon_0 \, n\,\omega_l^2\,m_e},
\end{equation}
with  $\omega_l$ the laser frequency, $n$ the refractive index of the semiconductor, $e$ the elementary charge and $m_e$ the effective mass of majority charge carriers. That is why the local charge carrier fluctuations $\eta$ cause fluctuations of the refractive index $\delta n$. These fluctuations are probed by any laser beam propagating through the optical element.

\begin{figure}
	\includegraphics[width=0.45\textwidth]{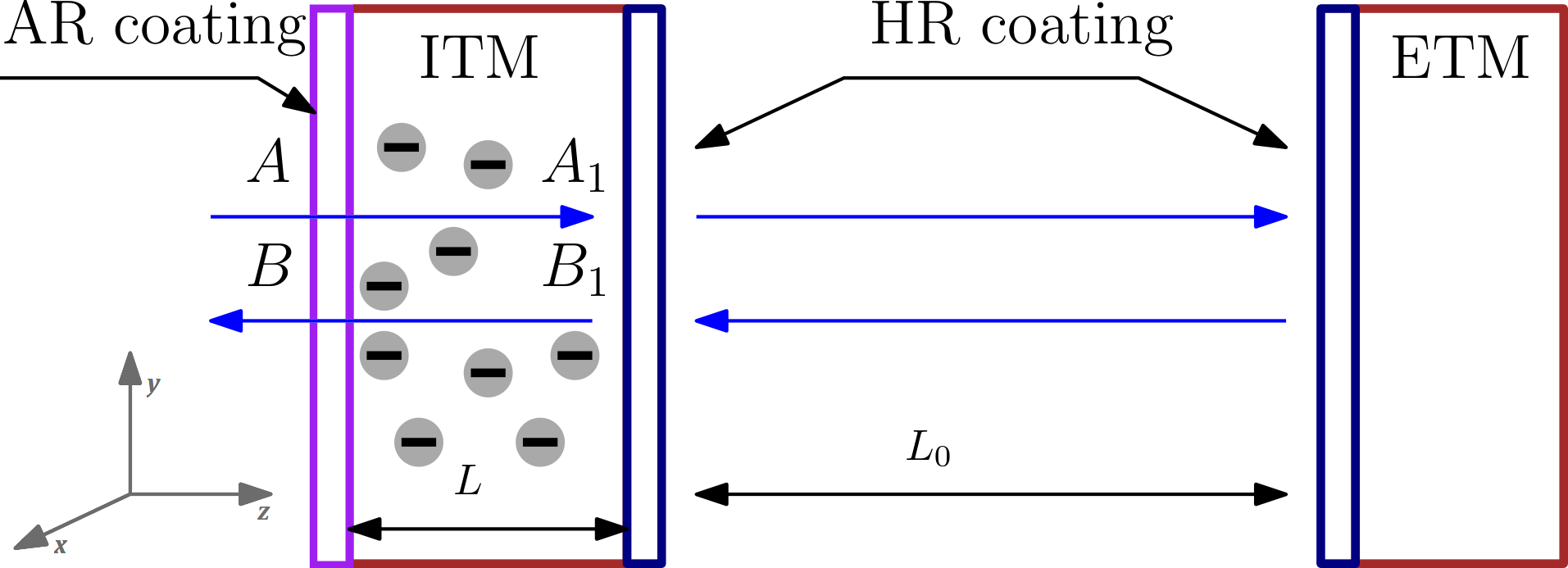}
	\caption{Fabry-Perot cavity. The surfaces of the input test mass are covered by anti-reflecting (AR) and high reflecting (HR) coatings. The thermal motion of carriers in the input test mass produces additional fluctuations of the refractive index wich are probed by the transmitted beams. The position of the coordinate system has been been shifted for better visibility. Its center is actually at the center of mass of the input mirror.}\label{FP}
\end{figure}

When a Gaussian laser beam propagates a distance $L$ along the $z$-axis inside of e.g. the input mirror (test mass) of a FP cavity (see Fig.~\ref{FP}), it reads out the information on the variation of the refractive index inside the input test mass weighted by the light intensity distribution in the cross section. Hence, the optical path length of the laser beam is shifted by a value of $\xi$:
\begin{align}
	\label{xi1}
	\xi(t) &=  \beta \int  \eta(\vec r, t)\cdot \Psi(\vec r)\, d\vec r, \\
	& \Psi(\vec r)  = \left\{\begin{array}{c c}
		\frac{e^{-(x^2+y^2)/r_0^2}}{\pi r_0^2},\quad& \text{if } 
		|z|\le \frac{L}{2}\\
		0, & \text{if } |z| > \frac{L}{2}
	\end{array}
	\right. ,
\end{align}
where $L$ is the length of the mirror substrate,  $r_0$ is the Gaussian radius of the light beam (referring to the length where the amplitude falls off by ${e^{-1}}$), $x,\, y$ are the transversal coordinates and $\int d\vec r$ is an integration over the volume of the input test mass. In the measurement we cannot distinguish between a fluctuation of the optical path length caused by noise or a length shift of the mirror caused by a gravitational wave. In this way carrier density fluctuations $\eta$ inside transmissive semiconductor optics contribute to the measured strain as $h_\xi$:
\begin{equation*}
	\eta \rightarrow \xi \rightarrow h_\xi.
\end{equation*}
We call this noise thermal charge carrier refractive (TCCR) noise. In order to quantify TCCR noise we have to model the carrier density fluctuations under consideration of screening effects that occur for free charge carriers (section \ref{secB}). Furthermore we have to understand how a change in the optical path length $\xi$ is related to the strain $h$ caused by the gravitational wave (section \ref{secA}) and combine both to get an expression for the amplitude of TCCR noise (section \ref{secC}). Finally we use this to explicitly compute the TCCR noise of ET input test masses (section \ref{sec3}). 

\subsection{Coupling of noise induced changes in the path length and the measured strain}\label{secA}
For simplicity reasons we consider a Michelson configuration with interferometer arms perpendicular to each other. To establish our model for TCCR noise we examine a Fabry-Perot (FP) cavity with mirrors as free test masses as shown in Fig.~\ref{FP}. 
TCCR noise in the input test mass disturbs the optical path length $n L$ by a value of 
$\xi$. From input-output relations we can see that the reflected amplitude $B$ (see notations on Fig.~\ref{FP}) contains an additional fluctuation term in the frequency domain:
\begin{align}
	\label{B}
	b_\xi &=  \frac{\gamma+i\Omega}{\gamma- i\Omega}\cdot A ik\xi + A ik\xi=
	Aik\xi\cdot \frac{2\gamma}{\gamma -i\Omega},
\end{align}
where $\Omega$ is a spectral frequency, $\gamma= T^2/2\tau$ is the amplitude relaxation rate of the FP cavity, $\tau=2L_0/c$ is the round trip time (usually $\Omega\tau\ll 1$) and  $T$ and $R$ are the amplitude transmittance and reflectance of the input test masses ($R^2+T^2=1$, we assume $T\ll 1$). The two terms in the first equation of \eqref{B} describe the perturbation created in the incident wave (after circulation inside the cavity) and the directly reflected one respectively. Effectively the cavity acts as a low pass Lorentz filter.


The fluctuation term should be compared with the signal term $b_s$ produced by the displacement $\delta L= L_0h_s/2$ ($h_s$ is the perturbation of the metric caused by the gravitational wave):
\begin{align}
	\label{b}
	b_s=T\cdot \frac{2 A}{T} \cdot \frac{2ik\, \delta L}{1-Re^{i\Omega\tau}}
	=\frac{2 A \cdot ik L_0}{\tau(\gamma-i\Omega)}\cdot h_s.
\end{align}
In gravitational wave detectors of the third generation there are  FP cavities in the east and north arms, the light from them interfere on a 50/50 beam splitter. Thus, the
signal $b_s$ \eqref{b} of the gravitational wave is increased by a factor of~$\sqrt 2$:
\begin{align}
	\label{bET}
	b_s^\text{ET} \ \Rightarrow\ b_s\sqrt 2.
\end{align}
On the other hand there are fluctuations $\xi_n$ and $\xi_e$ in the input test masses of the north and east arm cavities. As follows:
\begin{align} 
	b_\xi^\text{ET}= Aik\cdot \frac{2\gamma}{\gamma -i\Omega}\cdot \xi_\text{ET}, \quad
	\xi_\text{ET}=\frac{\xi_e +\xi_n}{\sqrt 2}. 
	\label{BET}
\end{align}
Note that the fluctuations $\xi_e$ and $\xi_n$ do not depend on each other and have equal power spectral densities. The power spectral density of $\xi_{ET}$ is equal to the power spectral density of $\xi_e$ (or $\xi_n$).
Comparing \eqref{B} and \eqref{b} by taking into account \eqref{bET} and \eqref{BET} one can find that the variation of the gravitational metric $h$ and the optical path fluctuations $\xi$ are related as:
\begin{align}
	\label{h}
	h = h_s +  h_\xi, \quad h_{\xi}=\frac{\gamma\tau}{\sqrt 2}\cdot \frac{\xi_{ET}}{L_0}=\frac{\pi}{\sqrt 2\,{F}}\cdot \frac{\xi_{ET}}{L_0},
\end{align}
with $F$ the Finesse of the Fabry-Perot cavity.

\subsection{Thermal charge carrier density fluctuations}\label{secB}
As suggested by the fluctuation-dissipation theorem we have to understand the dissipation mechanism (in our case that is diffusion) to describe fluctuations of charge carriers. For this reason we derive a diffusion equation including screening effects. We expect Debye screening to suppress diffusion processes thus reducing the fluctutations in the optical elements. Because of the residual doping in the large test mass substrates required for GW detection a majority charge carrier is present \cite{Degallaix_2014}. Without loss of generality we assume electrons to be the majority charge carriers. We start with the continuity equation for the carrier concentration $n_{cc}$~\cite{Jackson}:
\begin{align}
	\label{eq1} 
	\frac{\partial n_{cc}}{\partial t} &= - \vec\nabla \cdot \vec J,
\end{align}
where $\vec J$ is the current density described by~\cite{Jackson}:
\begin{align}
	\label{eq2}
	\vec J &= -D \vec\nabla n_{cc} +n_{cc}  \mu\, \vec E,\\
	\mu &= \frac{eD}{k_BT}. 
\end{align}
Here $D$ is the diffusion coefficient, $\mu$ the carrier mobility, $e$ the carrier charge, $k_B$ is the Boltzmann constant and $T$ the temperature. The last term in \eqref{eq2}, which is proportional to a small electric field $\vec E$, is introduced to account for Debye screening. The carrier concentration $n_{cc}$ can be described as a sum of a large constant~$n_0$ and a small fluctuating part~$\eta$: 
\begin{equation}
\label{n}
n_{cc}= n_0  + \eta,\quad n_0= \frac{1}{e \mu\rho_{e}}\,,
\end{equation}
with the electrical resistivity $\rho_e$. 
The screening electric field $\vec E$ is given by the Poisson equation~\cite{Jackson}: 
\begin{align}
	\label{E}
	\vec \nabla \cdot \vec E =\frac{e n_{cc}}{\epsilon \epsilon_0} \simeq \frac{e \eta}{\epsilon \epsilon_0} .
\end{align}
Here $\epsilon = n^2 $ with $n$ the refractive index and $\epsilon_0$ is the vacuum permittivity. 
After substituting \eqref{eq2} into \eqref{eq1} whilst using \eqref{n} we get:
\begin{align}
	\label{eq3c}
	\frac{\partial \eta}{\partial t} 
	& =  D \Delta \eta  - n_0 \mu\,\vec\nabla\cdot \vec E.
\end{align}
Since $\eta$ is a small deviation from thermal equilibrium (i.e. a fluctuation) we have neglected any quadratic term of $\eta$. Finally we get a diffusion equation utilizing \eqref{E}:
\begin{align}
	\label{eq3}
	\frac{\partial \eta}{\partial t} &= D\left(\Delta\, \eta - \frac{ 1}{\ell_D^2} \eta\right),\quad 
	\ell_D=\sqrt \frac{\epsilon\epsilon_0 k_BT}{n_0 e^2}\,.
\end{align}
Here $\Delta=(\vec\nabla)^2$ is the Laplace operator and $ \ell_D$ is the Debye length. Compared to conventional diffusion equations there is an additional term, $-D\,\eta/\ell_D^2$, which is a direct consequence of screening effects under the assumption of only having small deviations from thermal equilibrium $\eta$.

Next we want to include fluctuations. In the frame of Langevin approach \cite{vanVliet1980, braginsky1999} we introduce the Langevin forces $F (\vec r,t)$ into the right part of \eqref{eq3}:
\begin{align}
	\label{eq4}
	\frac{\partial \eta}{\partial t} = D\left(\Delta\, \eta - \frac{ 1}{\ell_D^2} \eta\right) + F (\vec r,t).
\end{align}
These Langevin forces $F (\vec r,t)$ cause the Brownian movement of the charge carriers. They are uncorrelated representing white Gaussian noise:
\begin{equation}
\begin{aligned}
	\label{eq4b}
	\langle F(\vec k,\omega) F^*(\vec k',\omega')\rangle =(&2\pi)^4 F_0^2 \left({k^2+ \ell_D^{-2}}\right)\\
	&\times \delta(\vec k -\vec k')\, \delta(\omega-\omega').
\end{aligned}
\end{equation} 
The correlators of the fluctuation forces and the constant $F_0$ should be chosen to fulfil the equation:
\begin{align}
	\label{n12}
	\left\langle \eta^2 \right\rangle= \frac{n_0}{V},
\end{align}
which corresponds to the well known requirement (of a Poisson process) that the variation $\langle\Delta N^2\rangle$ of the particle number in a volume $V$ is equal to the mean particle number: $\langle\Delta N^2\rangle = \langle N\rangle$~\cite{LL5}.  From \eqref{n12} one can find:
\begin{align}
	\label{F0}
	F_0^2=2 Dn_0.
\end{align}
Now we can calculate the auto-correlation function of density fluctuations:
\begin{equation}
\begin{aligned}
	B(\vec \rho,\tau)&\equiv \left\langle \eta(\vec r, t) \eta(\vec r- \vec \rho,t-\tau) \right\rangle=\\
	& = \frac{2Dn_0}{(2\pi)^4}  \int d\vec k\, d\omega \, 
	\frac{\left(k^2+\frac{1}{\ell_D^2}\right)e^{i\omega \tau+i\vec k \vec \rho}}{D^2 \left(k^2 + \frac{1}{\ell_D^2}\right)^2 +\omega^2}.
\end{aligned}
\end{equation}
From that the {\em double-sided}  power spectral density (PSD) $S_{\eta,\eta}(\vec k,\omega)$ can be found using the Wiener-Khinchin theorem~\cite{chintchin1934}:
\begin{equation}
\begin{aligned}
	S_{\eta\eta}(\vec k,\omega) 
	&= \int B(\vec \rho,\tau)\,
	e^{-i\omega \tau -i\vec k \vec \rho}\,d\vec \rho\, d\tau=\\
	\label{etaeta3}
	& =n_0\cdot
	\frac{ 2D\left(k^2+\frac{1}{\ell_D^2}\right)}{D^2 \left(k^2 + \frac{1}{\ell_D^2}\right)^2 +\omega^2}.
\end{aligned}
\end{equation}

\subsection{Variations of the optical path length}\label{secC}
Now that we know the spectral noise density of charge carrier fluctuation \eqref{etaeta3} and how a shift of the optical change of path length relates to the gravitational strain \eqref{h} we can combine these two results with the readout of the Gaussian beam \eqref{xi1} to get the measurable TCCR noise in a gravitational wave detector.
The readout of the Gaussian beam, equation \eqref{xi1}, can be rewritten as: 
\begin{align}
	\xi(t)	&  = \beta \int \frac{d\vec k\, d\omega}{(2\pi)^4}\ \eta(\vec k, \omega)\ \Psi(\vec k)\
	e^{ i\omega t},
\end{align}
where $ \eta(\vec k, \omega)$ and $\Psi(\vec k)$ are Fourier transforms of $\eta(\vec r, t)$ and $\Psi(\vec r)$ respectively.
To get the PSD of $\xi$ we calculate the auto-correlation function of $\xi$:
	\begin{align}
		\label{xi3}
		\langle \xi(t)\, &\xi(t-\tau)\rangle =\beta^2
		\int \frac{d\vec k\, d\omega}{(2\pi)^4}\, \frac{d\vec k'\, d\omega'}{(2\pi)^4}\\
		&\times  \left\langle\eta(\vec k, \omega)\ \eta^*(\vec k', \omega') \right\rangle
		\Psi(\vec k)\ \Psi^*(\vec k')\
		e^{ i\omega t- i\omega' (t-\tau)}.\nonumber
	\end{align}
Using \eqref{eq4b} we can express the correlator as:
	\begin{align}
		\langle\eta(\vec k, \omega)\ &\eta^*(\vec k', \omega') \rangle =
		\frac{\left(k^2+\frac{1}{\ell_D^2}\right)}{
			\omega^2 + D^2\left(k^2+\frac{1}{\ell_D^2}\right)^2}\nonumber\\
		&\times (2\pi)^4\, F_0^2
		\cdot \delta(\vec k -\vec k')\, \delta(\omega-\omega'),
	\end{align}
and substituting into \eqref{xi3} and using the Wiener-Khinchin theorem and the normalisation \eqref{F0} we write down the double-sided PSD $S_{\xi\xi}(\omega)$:
\begin{align}
	\label{xixi}
	S_{\xi\xi}(\omega) &=  2Dn_0\beta^2
	\int \frac{d\vec k}{(2\pi)^3}\cdot
	\frac{\left(k^2+\frac{1}{\ell_D^2}\right) \big|\Psi(\vec k)\big|^2}{\omega^2 + D^2\left(k^2+\frac{1}{\ell_D^2}\right)^2}.
\end{align}
With the results of section \ref{secA} we can rewrite \eqref{xixi} in terms of the GW metric using \eqref{h}:
\begin{align}
	\label{xixih}
	S_{\xi\xi}^h(\omega) &=  \frac{n_0\beta^2(\gamma \tau)^2}{L_0^2}
	\int \frac{d\vec k}{(2\pi)^3}\cdot
	\frac{D\left(k^2+\frac{1}{\ell_D^2}\right) \big|\Psi(\vec k)\big|^2}{\omega^2 + D^2\left(k^2+\frac{1}{\ell_D^2}\right)^2}.
\end{align}
In the usual case, when the Debye length is much smaller than the physical dimensions of the systems, i.e. the Gaussian radius and the length of the substrate (see table \ref{tab:gianttable}), $\ell_D \ll r_0, L$ this expression can be simplified: 
\begin{align}
	\label{xixih2}
	S_{\xi\xi}^h(\omega) &=  \frac{n_0\beta^2(\gamma \tau)^2}{L_0^2}\cdot
	\frac{L}{2\pi\, r_0^2}\cdot
	\frac{ \tau_m }{1+\omega^2\tau_m^2 },\\
	& \tau_m = \frac{\ell_D^2}{D}= \epsilon\epsilon_0 \rho,
	\label{maxwelltime}
\end{align}
where $ \tau_m$ is the Maxwell relaxation time. Here we used the normalisation:
\begin{align}
	& \int \frac{d\vec k}{(2\pi)^3} \big|\Psi(\vec k)\big|^2 
	= \int d\vec r\, \big|\Psi(\vec r)\big|^2={ \frac{ L}{2 \pi\, r_0^2}}.
\end{align}
The PSD can be converted into the amplitude spectral density (ASD):
\begin{align}
	\label{xih}
	S_{\xi}^h(\omega)=\sqrt{S_{\xi\xi}^h(\omega)}\, .
\end{align}
Note that formula \eqref{xixih} for the spectral noise density is valid for frequencies smaller than the inverse diffusion time because the equation of motion \eqref{eq3c} does not account for retardation effects. The diffusion time is either determined by the Maxwell rexalation time or the electron lifetime whatever is smaller~\cite{Forster}. In silicon which is an indirect semiconductor the diffusion time is given by the Maxwell relaxation time \eqref{maxwelltime} which is around $10^{-8}$~s for pure silicon at room temperature. The limiting frequency is therefore around $10^8$~Hz which is much larger than the frequency range of third generation gravitational wave detectors of $1$~to~$10^4$ Hz~\cite{ETdesign}.

\section{Computation of TCCR noise}\label{sec3}

To quantify the magnitude of TCCR noise we are going to use the low-frequency interferometer of the Einstein telescope~\cite{ETdesign} as an example for a cryogenically operated third generation gravitational wave detector where the use of crystalline substrate materials is considered. Since we are examining transmissive TCCR noise the HR and AR coatings do not contribute significantly because of their small thickness. In amorphous coatings TCCR noise is even further reduced due to the low amount of free charge carriers.

\begin{figure}
	\includegraphics[width=0.5\textwidth]{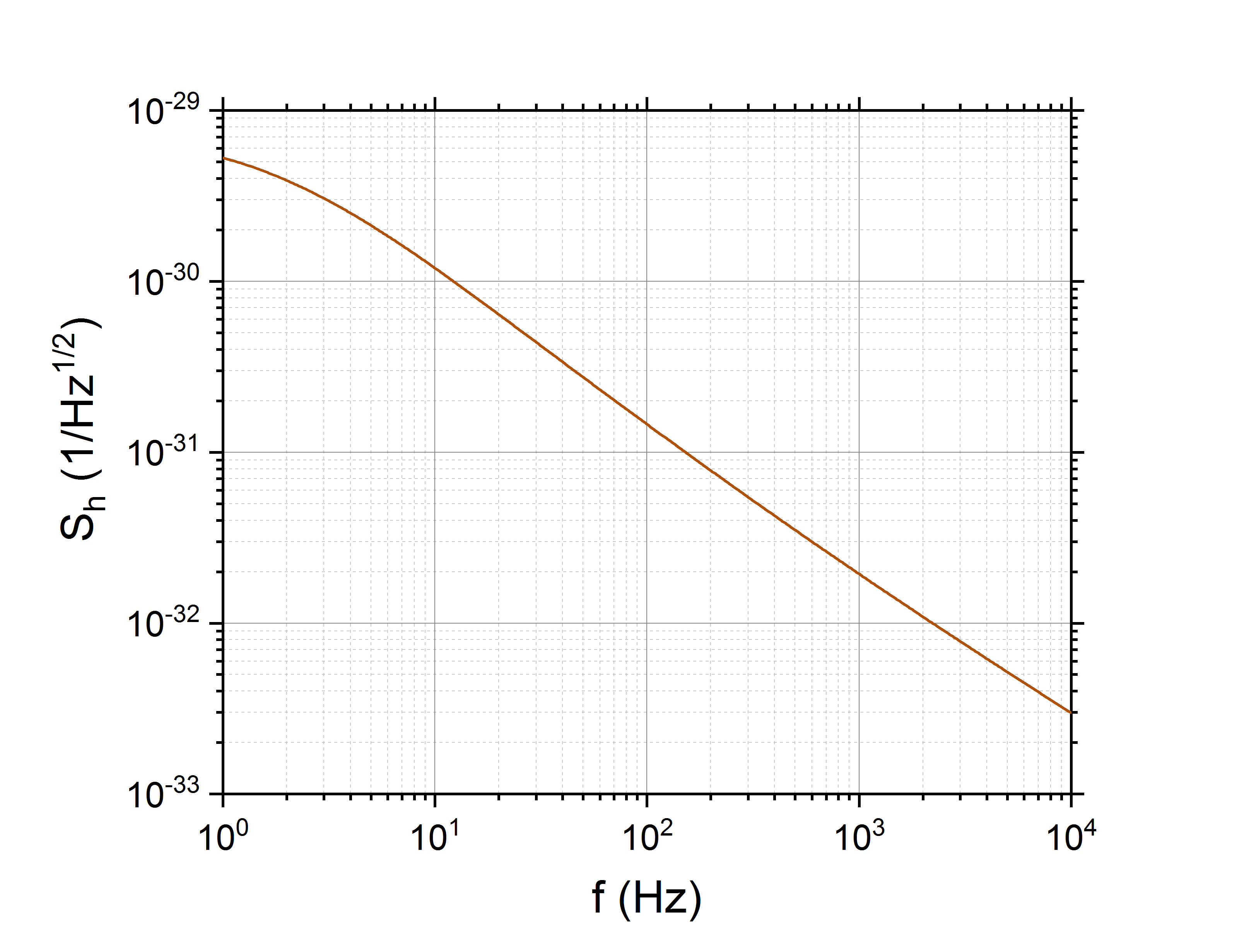} 
	\caption{ASD of TCCR noise in the input test mass substrates of ET at $10$~K for pure Si, $n_D=5\cdot 10^{12}$~$\frac{1}{\rm{cm}^3}$. The ASD falls off with $f^{-1}$ akin to a Brownian noise source. The frequency independent, approximated formula \eqref{xixih2} cannot be used in this case because of the freeze out of charge carriers enabling the Debye length to become of order of the Gaussian beam radius (see table \ref{tab:gianttable}).}\label{fig:10K}
\end{figure}

\begin{figure}
	\includegraphics[width=0.5\textwidth]{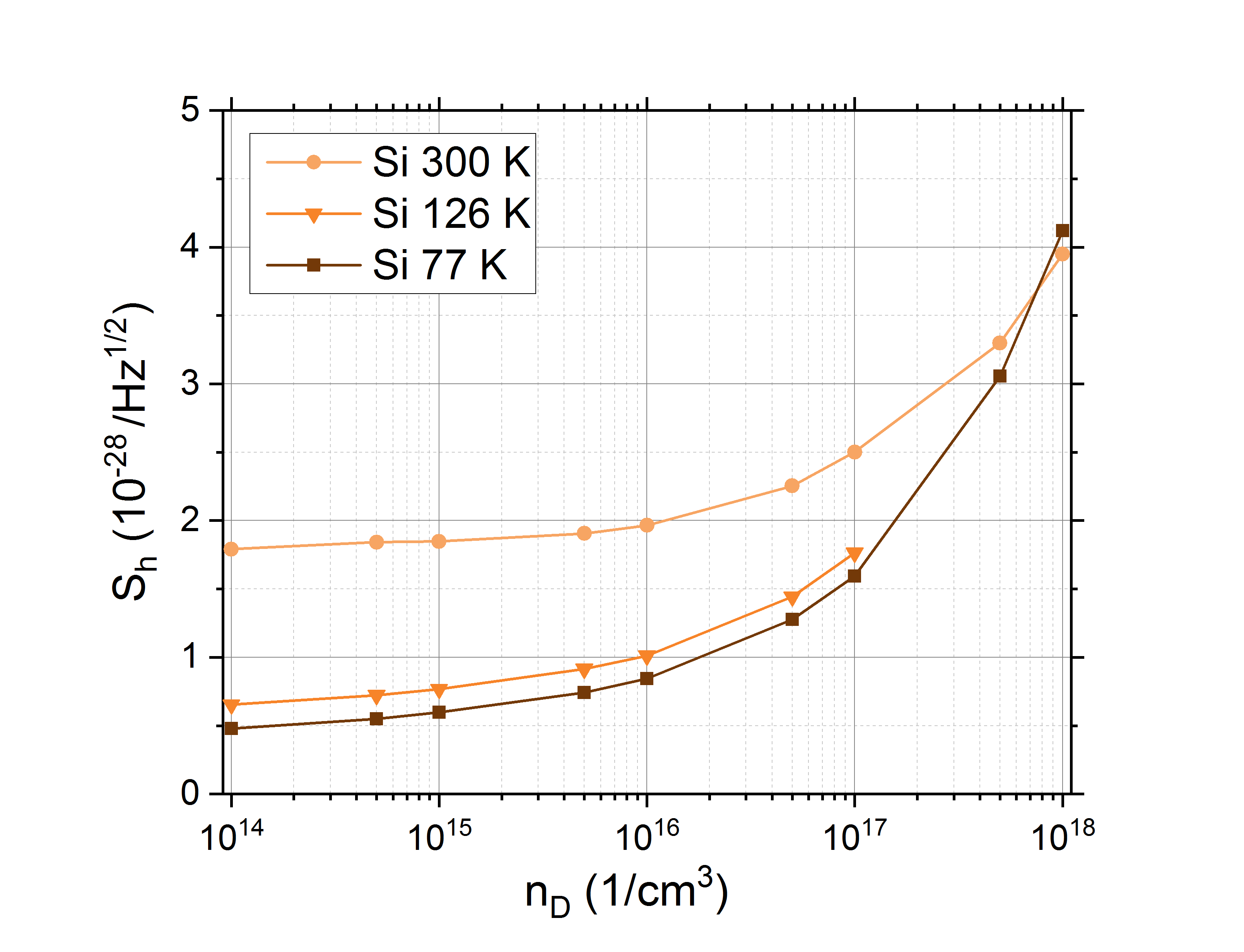}
	\caption{Doping dependency of the ASD of TCCR noise $S^\xi_h$. At the shown temperatures the ASD is a constant of frequency in the frequency interval of $1$ to $10^{4}$~Hz  for Si. Due to a lack of data for the charge carrier mobility $\mu$ of doped materials at cryogenic temperatures we show the temperatures $77$~K, $126$~K and $300$~K. The temperature of $126$~K is interesting because at this temperature the thermoelastic coefficient of Si vanishes immensely reducing the thermoelastic noise. The ASD of Si at $77$~K and $300$~K cross due to a crossing of the charge carrier mobility at high levels of doping (see table \ref{tab:gianttable}).}\label{fig:doping}
\end{figure}

\begin{table}[tb]
   \caption{Parameters of the input test masses of the Einstein telescope~\cite[p.347]{ETdesign}.}
		\label{tab:ET-Input}
	\begin{center}
		\begin{tabular}{l|c|r}
			parameter&symbol&value\\
			\hline
			finesse of FP cavity &$F$&$880$\\
			mirror thickness &$L$&$0.5$ m\\
			Gaussian radius &$r_0$&$0.09$ m\\
			laser wavelength &$\lambda_L$&$1550$ nm\\
			arm length &$L_0$&$10^4$ m\\
			refractive index Si~\cite{nsi}&$n$&$3.487$\\
			effective electron mass Si &$m_e$& $0.260$~$m_0$\\
			electron rest mass &$m_0$& $9.109\cdot10^{-31}$ kg\\
			Boltzmann constant& $k_B$ & $1.380\cdot10^{-23}$ $\frac{\rm{J}}{\rm{K}}$\\
			vaccum permittivity & $\epsilon_0$ & $8.854 \cdot 10^{-12}$ $\frac{\rm{AS}}{\rm{Vm}}$
		\end{tabular}
		
	\end{center}
\end{table}

TCCR noise is dependent on the diffusion coefficient $D$ and the Debye length $\ell_D$ (see table \ref{tab:gianttable} for numerical values of $D$ and $\ell_D$). This is why it is indirectly also a function of temperature $T$ and doping density $n_D$. At this time large high-purity silicon samples can be produced with the magnetically assisted Czochralski technique with a resistivity of up to $10$~k$\Omega$cm. To account for production difficulties we are looking at n-type silicon with a worse resistivity of $1$~k$\Omega$cm which corresponds to a doping density of around $4.4\cdot 10^{12}$~$\frac{1}{\rm{cm}^3}$, which is equivalent to a resistivity of $1\,$k$\Omega$cm at room temperature~\cite{Degallaix_2014}.
In order to calculate the mean carrier density $n_0$ as a function of the doping density a model for semiconductors with no compensation (no acceptors) has been used~\cite{Grundmann}:
\begin{align}
	n_0=\frac{2n_D}{1+\left[1+4g_D\,\frac{n_D}{n_C}\,\textrm{exp}\left(\frac{E_D^\textrm{b}}{k_B T}\right)\right]^{1/2}}.
\end{align}
Here $g_D$ is the donor degeneracy, $n_D$ is the donor concentration, $E_D^b$ is the energy gap between donor level and the conduction band and $n_c$ is the conduction band edge density:
\begin{align}
	\label{edgedensity}
	n_c=2\left(\frac{m_{e,b}^*\,k_BT}{2\pi\hbar^2}\right)^{{3}/{2}},
\end{align}
with $m_{d,e}^*$ the effective density-of-states mass of the conduction band electrons. As dopants we considered shallow phosphorous donors which have a degeneracy of $g_D=2$ and an energy gap of $E_D^b=45$~meV.

Additionally, we want to be able to model the diffusion coefficient $D$. For this we used the Einstein-Smoluchowski equation in order to replace the diffusion coefficient $D$ with the well investigated charge carrier mobility $\mu$~\cite{Grundmann}:
\begin{align}
	D=\frac{k_B T}{e} \mu.
\end{align}
A more precise expression for $D$ can be obtained from taking into account the carrier band curvature. In our case though, the Fermi level is several $eV$ deep within the band gap. Thus,  the deviations from the Einstein-Smoluchowski equation are negligibly small. The mobility data for modeling $D$ has been obtained from Jacoboni \textit{et al.}~\cite{jacoboni1977} and Li \textit{et al.}~\cite{li1977}.

In Fig. \ref{fig:10K} we have plotted the ASD of TCCR noise of the ET input test masses at its operation temperature of $10$~K for pure Si, $n_D\approx5\cdot10^{12}$~$\frac{1}{\rm{cm}^3}$ using the parameters shown in table \ref{tab:ET-Input} and \ref{tab:gianttable}. All numerical calculations shown have been performed with \textit{Mathematica}~11.3~\cite{mathematica113}. We can see that the ASD of TCCR noise at $10$~K reaches  $5.3\cdot10^{-30}$~$/\sqrt{\textrm{Hz}}$ in Si. This noise amplitude is well below the sensitivity goal of the ET telescope of $3\cdot10^{-25}$~$\sqrt{\textrm{Hz}}$. Therefore we can come to the important conclusion that TCCR noise is not a limiting noise source for the ET telescope and third generation gravitational wave detection in general.


Like all thermal noise sources TCCR noise scales with temperature, additionally it also increases with the doping concentration. In Fig.~\ref{fig:doping} we are showing the ASD of TCCR noise in dependence of the doping density for different temperatures. Because of the temperature range shown ($77$~K to $300$~K) the Debye length of Si (see table \ref{tab:gianttable}) is small enough to use the approximation given by formula \eqref{xixih2} thus obtaining a frequency independent ASD  up to at least $10^4$~Hz.

The doping dependence can be explained in the following way: An increase in the doping concentration increases the mean carrier concentration $n_0$ and decreases the Debye length $\ell_D$. These effects compensate each other as can be seen in equation \eqref{xixih2}. Furthermore in doped silicon ($n_D>10^{14}$~$\frac{1}{\textrm{cm}^3}$) an increase in the doping density $n_D$ also decreases the charge carrier mobility $\mu$~\cite{li1977} and the diffusion coefficient $D$ respectively (see table \ref{tab:gianttable}) thus increasing the TCCR noise amplitude. Overall even for room temperature (compared to cryogenic operation) and high doping densities TCCR noise stays below the sensitivity targets of third generation wave detectors. Calculations with other III-V semiconductors with a smaller band gap (such as gallium arsenide) have shown that the TCCR noise amplitude does not increase significantly.


\section{Conclusion}
 
 

We show a detailed analysis of thermal charge carrier refractive (TCCR) noise in metals and semiconductors from first principles especially taking into account screening effects. We show that Debye screening practically blocks diffusion processes, hence dramatically reducing TCCR noise. We apply our approach to the input masses of the Einstein telescope and show that TCCR noise is not a limiting noise source. Furthermore it can be seen that TCCR noise does not increase significantly at room temperature (compared to cryogenic operation) and high levels of doping of the material. For instance for silicon input test masses of ET at room temperature and a rather high doping density of $n_D=10^{16}$~$\frac{1}{\rm{cm}^3}$ the TCCR noise amplitude is three orders of magnitude below the design sensitivity. Before this paper the driver for ultra-pure silicon has been optical absorption and TCCR noise. With our derivation we show that TCCR noise is not a limiting noise source for substrate materials of third generation gravitational wave detectors. Nevertheless there is still a need for ultra-pure substrate materials due to limits set by the optical absorption~\cite{winkler1991}~\cite{Degallaix_2014}.

\acknowledgements
S.V. acknowledges support from  Russian Foundation of Basic Research (Grant No. 19-29-11003) and from the TAPIR GIFT MSU Support of California Institute of Technology.
S.K. and F.B. acknowledge support by the German Research Foundation (DFG) under Germany's Excellence Strategy EXC-2123 QuantumFrontiers - 390837967.
This document has LIGO number LIGO-P2000048.

\bibliographystyle{unsrt}
\bibliography{TD} 

\appendix

\begin{table*}[tb]
	\caption{Electron mobility $\mu$, diffusion coefficient $D$ and Debye length $\ell_D$ of Si at different temperatures. The mobility data of Si at $126$~K has been approximated by taking the mean of the mobilities at $100$~K and $150$~K from Li \textit{et al}~\cite{li1977}.}
	\label{tab:gianttable}
	\begin{center}
		\begin{tabular}{r|r||c|c|c}
			\multicolumn{2}{c||}{ }&\multicolumn{3}{c}{Silicon} \\
			\hline \hline
			$T [K]$ & $n_D$ [1/$\textrm{cm}^3$] & $\mu$ [$\textrm{m}^2$/Vs]& $D$ [$\textrm{m}^2$/s] & $\ell_D$ [m]\\
			\hline \hline
			$10$ & $5\cdot10^{12}$ & $38.8$ \cite{jacoboni1977} & $3.34\cdot10^{-2}$ & $9.38\cdot10^{-2}$ \\
			\hline
			$77$ & $10^{14}$ & $2.04$ \cite{li1977}& $1.35\cdot10^{-2}$ & $6.84\cdot10^{-7}$\\
			$77$ & $5\cdot10^{14} $ & $1.55$ \cite{li1977}& $1.03\cdot10^{-2}$ & $4.51\cdot10^{-7}$\\
			$77$ & $10^{15}$ & $1.31$ \cite{li1977}& $8.67\cdot10^{-3}$ & $3.78\cdot10^{-7}$\\
			$77$ & $5\cdot10^{15}$ & $0.845$ \cite{li1977}& $5.61\cdot10^{-3}$ & $2.52\cdot10^{-7}$\\
			$77$ & $10^{16}$ & $0.655$ \cite{li1977}& $4.35\cdot10^{-3}$ & $2.12\cdot10^{-7}$\\
			$77$ & $5\cdot10^{16}$ & $0.287$ \cite{li1977}& $1.90\cdot10^{-3}$ & $1.41\cdot10^{-7}$\\
			$77$ & $10^{17}$ & $0.184$ \cite{li1977}& $1.22\cdot10^{-3}$ & $1.19\cdot10^{-7}$\\
			$77$ & $5\cdot10^{17}$ & $0.050$ \cite{li1977}& $3.32\cdot10^{-4}$ & $7.94\cdot10^{-8}$\\
			$77$ & $10^{18}$ & $0.027$ \cite{li1977}& $1.82\cdot10^{-4}$ & $6.67\cdot10^{-8}$\\
			\hline
			$126$ & $10^{14}$ & $1.09$ & $1.19\cdot10^{-2}$ & $4.19\cdot10^{-7}$\\
			$126$ & $5\cdot10^{14}$ & $0.895$ & $9.72\cdot10^{-3}$ & $2.61\cdot10^{-7}$\\
			$126$ & $10^{15}$ & $0.795$ & $8.63\cdot10^{-3}$ & $2.15\cdot10^{-7}$\\
			$126$ & $5\cdot10^{15}$ & $0.557$ & $6.05\cdot10^{-3}$ & $1.41\cdot10^{-7}$\\
			$126$ & $10^{16}$ & $0.456$ & $4.95\cdot10^{-3}$ & $1.18\cdot10^{-7}$\\
			$126$ & $5\cdot10^{16}$ & $0.224$ & $2.43\cdot10^{-3}$ & $7.81\cdot10^{-8}$\\
			$126$ & $10^{17}$ & $0.150$ & $1.63\cdot10^{-3}$ & $6.55\cdot10^{-8}$\\
			\hline
			$300$ & $10^{14}$ & $0.146$ \cite{li1977}& $3.76\cdot10^{-3}$ & $4.33\cdot10^{-7}$\\
			$300$ & $5\cdot10^{14}$ & $0.138$ \cite{li1977}& $3.56\cdot10^{-3}$ & $2.14\cdot10^{-7}$\\
			$300$ & $10^{15}$ & $0.137$ \cite{li1977}& $3.54\cdot10^{-3}$ & $1.64\cdot10^{-7}$\\
			$300$ & $5\cdot10^{15}$ & $0.129$ \cite{li1977}& $3.32\cdot10^{-3}$ & $9.52\cdot10^{-8}$\\
			$300$ & $10^{16}$ & $0.121$ \cite{li1977}& $3.13\cdot10^{-3}$ & $7.73\cdot10^{-8}$\\
			$300$ & $5\cdot10^{16}$ & $0.092$ \cite{li1977}& $2.38\cdot10^{-3}$ & $4.93\cdot10^{-8}$\\
			$300$ & $10^{17}$ & $0.075$ \cite{li1977}& $1.93\cdot10^{-3}$ & $4.10\cdot10^{-8}$\\
			$300$ & $5\cdot10^{17}$ & $0.043$ \cite{li1977}& $1.11\cdot10^{-3}$ & $2.70\cdot10^{-8}$\\
			$300$ & $10^{18}$ & $0.030$ \cite{li1977}& $7.74\cdot10^{-4}$ & $2.26\cdot10^{-8}$\\
			
		\end{tabular}
		
	\end{center}
\end{table*}

\end{document}